\documentclass{article}
\usepackage[dvips]{graphicx}
\begin{document}
\title{Noise-induced synchronization for phase turbulence}
\author{Hidetsugu  Sakaguchi\\
Department of Applied Science for Electronics and Materials,\\ Interdisciplinary Graduate School of Engineering Sciences,\\
Kyushu University, Kasuga, Fukuoka 816-8580, Japan}
\maketitle
\begin{abstract}
Phase turbulence is suppressed  by applying common noise additively 
to the Kuramoto-Sivashinsky type equation, and the noise-induced phase
 synchronization is realized. 
The noise strength necessary for the suppression of phase turbulence is 
evaluated theoretically.
\end{abstract}
Recently, various noise effects to nonlinear systems have been studied.
The response of a bistable system or an excitable system to a periodic force is enhanced by the noise effect. 
The stochastic resonance improves signal detection by the superposed noise \cite{rf:1,rf:2,rf:3}.  
 Noise-enhanced entrainment among coupled oscillators is found experimentally in   Belousov-Zhabotinsky reactions \cite{rf:4}.
Frequency locking of noise-sustained oscillations is found in coupled excitable systems \cite{rf:5,rf:6}.
A small amount of noise may change a chaotic trajectory into a 
rather regular trajectory, and it is called noise-induced order \cite{rf:7}.
Common noise may induce complete synchronization for uncoupled chaotic oscillators \cite{rf:8,rf:9}. It is called noise-induced synchronization. 
In this paper, we apply common noise for a modified equation of the Kuramoto-Sivashinsky equation.  Without the common noise, the model equation exhibits phase turbulence.  Noise-induced synchronization occurs and a spatially uniform state is observed owing to the common noise.

The model equation is written as 
\begin{equation}
\phi_t=\omega-r\sin\phi-\mu \phi_{xx}-\phi_{xxxx}+\phi_x^2+\xi(t),
\end{equation}
where $\phi(x,t)$ is a phase variable, $\omega>r$ is assumed, and $\xi(t)$ represents spatially uniform Gaussian white noise satisfying 
\[\langle \xi(t)\xi(t^{\prime})\rangle=2D\delta(t-t^{\prime}).\]
The model equation is equal to the original Kuramoto-Sivashinsky equation, when  
$r=D=0$.  
The Kuramoto-Sivashinsky equation exhibits phase turbulence for $\mu>0$, owing to the long-wavelength instability of the uniform state. 
For nonzero $r$, the phase motion of each oscillator is not 
uniform.  This type of phase oscillator model has been used to study cooperative phenomema in coupled limit cycle oscillators \cite{rf:5, rf:10,rf:11}.
Equation (1) represents a model of spatially coupled phase oscillators. Without the noise term, the spatially uniform rotation is unstable and 
a phase turbulence is observed similarly to the original Kuramoto-Sivashinsky equation. 

\begin{figure}[htb]
\begin{center}
\includegraphics[width=12cm]{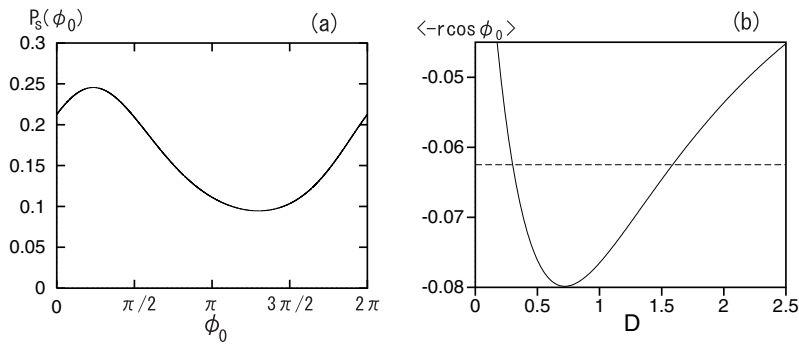}
\caption{(a) Stationary probability distribution $P_s(\phi_0)$ for the Fokker-Planck equation at $\omega=0.8,\;r=0.5$ and $D=0.7$. (b) Average value of $-r\cos\phi_0$ as a function of $D$. Below the dashed line $\langle -r\cos\phi_0\rangle=-0.0625$, the uniform state is stable for $\omega=0.8,\;r=0.5$ and $\mu=0.5$.  
}
\label{fig:1} 
\end{center}
\end{figure} 
\begin{figure}[htb]
\begin{center}
\includegraphics[width=11cm]{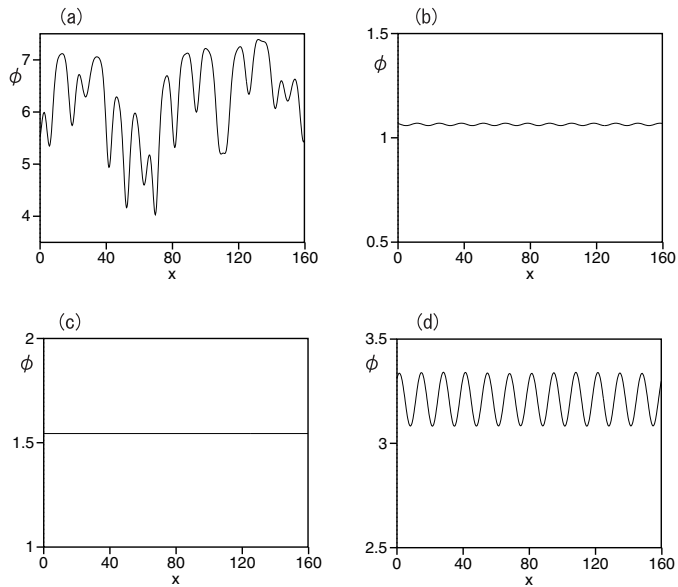}
\caption{Snapshot profiles $\phi(x,t)$ for Eq.~(1) at $\omega=0.8,\;r=0.5$ and $D=0$ (a), 0.2 (b), 0.7 (c) and 2.5 (d).
}
\label{fig:2} 
\end{center}
\end{figure} 
\begin{figure}[htb]
\begin{center}
\includegraphics[width=11cm]{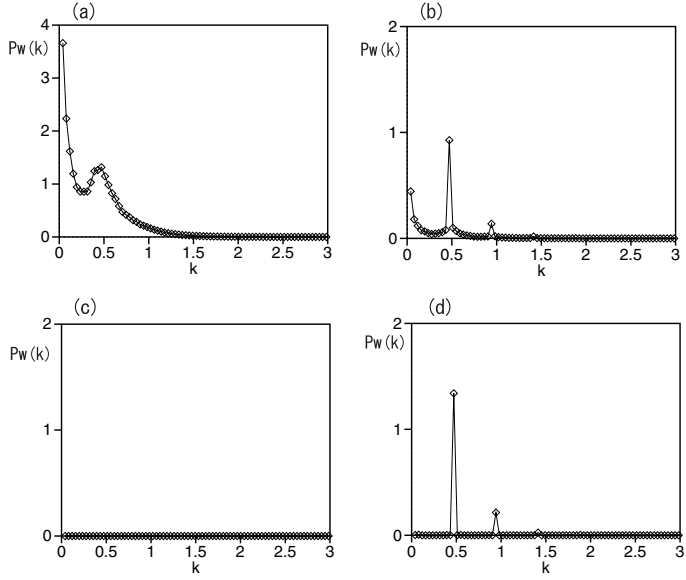}
\caption{Average of Fourier amplitude $Pw(k)=\langle |\phi_k|\rangle$ at $\omega=0.8,\;r=0.5$ and $D=0$ (a), 0.2 (b), 0.7 (c) and 2.5 (d).  
}
\label{fig:3} 
\end{center}
\end{figure} 
\begin{figure}[htb]
\begin{center}
\includegraphics[width=11cm]{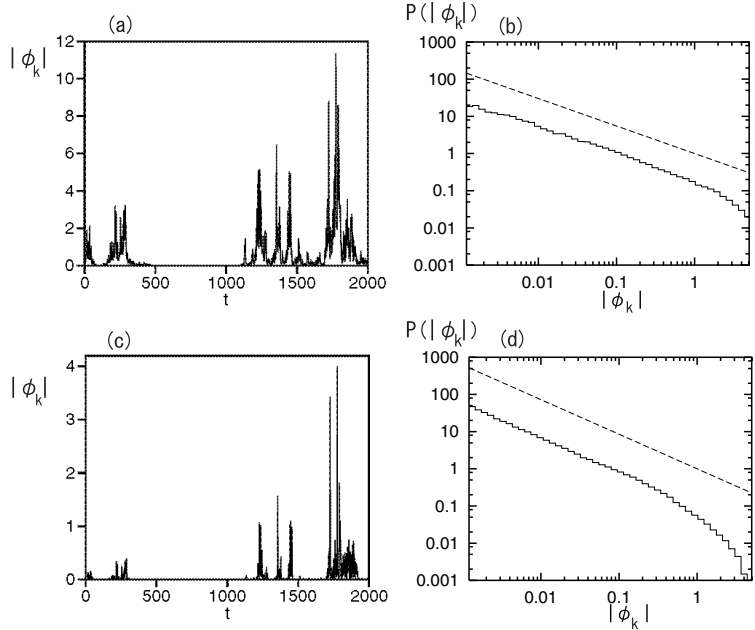}
\caption{(a) Intermittent time sequence of a Fourier amplitude $|\phi_{k}(t)|$ with wavenumebr $k=0.471$ for $\omega=0.8,\;r=0.5$ and $D=0.2$.
(b) Histograms of the probability distributions of $|\phi_k|$ with wavenumber $k=0.471$. The straight line denotes a power law with exponent -0.74. (c) Intermittent time sequence of a Fourier amplitude $|\phi_{k}(t)|$ with wavenumebr $k=0.942$. 
(d) Histograms of the probability distributions of $|\phi_k|$ with wavenumber $k=0.942$. The straight line denotes a power law with exponent -0.93.}
\label{fig:4} 
\end{center}
\end{figure} 

Firstly, we study the stability of the spatially uniform state under the common noise.  The spatially uniform state $\phi(x,t)=\phi_0(t)$ satisfies the Langevin equation 
\begin{equation}
d\phi_{0}/dt=\omega-r\sin\phi_0+\xi(t).
\end{equation}
The corresponding Fokker-Planck equation is 
\begin{equation}
\frac{\partial P}{\partial t}=-\frac{\partial}{\partial \phi_0}\{(\omega-r\sin\phi_0)P\}+D\frac{\partial^2 P}{\partial \phi_0^2}.
\end{equation} 
For $D=0$, the stationary probability distribution is
\begin{equation}
P_s(\phi_0)\propto \frac{1}{\omega-r\sin\phi_0}.
\end{equation}
For nonzero D, the stationary probability distribution is expressed as \cite{rf:11,rf:12}
\begin{equation}
P_s(\phi_0)=\exp\{(\omega \phi_0-r+r\cos\phi_0)/D\}P_s(0)\left\{1+\frac{(e^{-2\pi\omega/D}-1)\int_0^{\phi_0} e^{(-\omega\psi-r\cos\psi)/D}d\psi}{\int_0^{2\pi}e^{(-\omega\psi-r\cos\psi)/D}d\psi}\right \},
\end{equation}
where $P_s(0)$ is determined from the normalization condition $\int_0^{2\pi}P_s(\phi_0)d\phi_0=1$.
This stationary probabiliy distribution approaches a uniform distribution $P_s=1/2\pi$ for  $D\rightarrow \infty$. 
The stationary distribution for $\omega=0.8,\;r=0.5$ and $D=0.7$ is displayed in Fig.~1(a).  The peak of the probability distribution is located at $\phi\sim \pi/4<\pi/2$.

The linearized equation of the Fourier amplitude $\phi_k$ with wave number $k$  around the uniform state for Eq.~(1) satisfies an equation
\begin{equation}
d\phi_k/dt=(-r\cos\phi_0(t)+\mu k^2-k^4)\phi_k,
\end{equation}
where $\phi_k=1/\sqrt{L}\int_0^L\phi(x)e^{-ikx}dx$, and $\phi_0(t)$ obeys Eq.~(2). 
The long-time average of $-r\cos\phi_0(t)+\mu k^2-k^4$ 
determines the stability for the perturbation with wave number $k$. 
For $D=0$, the average value of $\cos\phi_0$ is zero, since the distribution (4) is symmetric around $\phi=\pi/2$. 
For $D\rightarrow \infty$, the average value of $\cos\phi_0(t)$ is also zero, since the distribution is uniform.  For general parameter values of $D$,  the average value of $-r\cos\phi_0(t)$ is negative, since the probability distribution has a peak below $\pi/2$ as shown in Fig.~1(a), and $\cos\phi_0$ is positive for $0<\phi_0<\pi/2$.  We have calculated the average value of $-r\cos\phi_0(t)$ using the stationary distribution (5) for various $D$'s for $\omega=0.8$ and $r=0.5$, and plotted the results in Fig.~1(b). 
The curve of $\langle -r\cos\phi_0\rangle$ has a characteristic form of the resonant one, although it is not easily expected from the distribution (5).  The average value of $-r\cos\phi_0$ approaches 0 for $D\rightarrow 0$ and $D\rightarrow \infty$ and it takes a minimum value -0.08 at $D\sim 0.7$.
 The average value of $-r\cos\phi_0(t)+\mu k^2-k^4$ takes a maximum $-r\langle \cos\phi_0\rangle+\mu^2/4$ at $k=\sqrt{\mu/2}$, when $k$ is changed. 
The uniform state is linearly stable if $-r\langle \cos\phi_0\rangle+\mu^2/4<0$. At $\omega=0.8$, $r=0.5$ and $D=0.7$, the uniform state is stable if $\mu$ is smaller than 0.4. For $\mu=0.5$, $\mu k^2-k^4$ takes a maximum value 0.0625 at $k=0.5$, therefore, the uniform state is stable for $0.3<D<1.58$. 
It implies that the common noise of intemediate strength can suppress long-wavelength fluctuations and stabilize the uniform state.  Common noise of intermediate strength is most effective for the noise-induced synchronization, which is analogous to the stochastic resonance.  

We have performed numerical simulation to see the noise-induced 
synchronization for the modified Kuramoto-Sivashinsky equation.
We have used the finite difference method with time step 0.0001 and space step 160/512 for a numerical simulation, and the periodic boundary conditions 
with system size 160 is used. The parameter values are $\omega=0.8$, $r=0.5$ and $\mu=0.5$.  The noise intensity $D$ was changed as a control parameter. 
Figure 2 displays snapshot profiles of $\phi(x)$ at $D=0, 0.2, 0.7 $ and 2.5. 
At $D=0$, the spatio-temporal chaos is observed. 
At $D=0.7$, the spatially uniform state appears. 
It corresponds to the noise-induced synchronization.
At $D=0.2$ and $D=2.5$, the spatial profile is not uniform, but fairly  
regular. The wavelength of the most dominant Fourier mode is 160/12, and the corresponding wavenumber is 0.471, which is close to $\sqrt{\mu/2}=0.5$. 
The amplitude of spatial fluctuations changes intermittently in time. 
The intermittent time sequence of the Fourier amplitude $|\phi_k(t)|$ with $k=0.471$ for $D=0.2$ is shown in Fig.~4(a).  
The amplitudes of the spatial fluctuations seem to differ very much for $D=0.2$ and 2.5 in Figs.~2(b) and (d), but it is a problem of timing to take snapshots.  The average amplitudes of the spatial fluctuations do not differ so much for the two parameter values of $D$, as shown in Figs.~3(b) and (d).

Figure 3 displays the average of the Fourier amplitude: $P_w(k)=\langle |\phi_k| \rangle$ for the four parameters of the noise intensity.  For $D=0$, the spectrum has a peak at $k\sim \sqrt{\mu/2}$ and increases near $k\sim 0$. This is a characteristic spectrum of 
phase turbulence. For $D=0.7$, the uniform state is stable and the spectrum is almost zero.  For $D=0.2$ and $D=2.5$, the spectrum has strong peaks at $k\sim \sqrt{\mu/2}$ and its higher harmonics, but the long-wavelength fluctuations are suppressed in contrast to the case of $D=0$. 
The suppression of the long-wavelength fluctuations makes the spatial profiles of $\phi(x)$ regular, as shown in Fig.~2.

The time evolution is intermittent, and  $\phi(x,t)$ is close to the uniform state $\phi_0(t)$, when the amplitude of spatial fluctuations becomes sufficiently small. Then, the Fourier amplitudes obey Eqs.~(2) and (6) approximately.   
The variable $\phi_0(t)$ is a random varibale, since Eq.~(2) is a Langevin equation. Equation (6) represents a random multiplicative process, since 
the growth rate $-r\cos\phi_0(t)+\mu k^2-k^4$ for the Fourier amplitude with wave number $k$ is randomly fluctuating. If the average value of the growth rate is positive (negative), the Fourier amplitude generally tends to grow from zero (decay to zero).  Even if the average value of the growth rate is positive, there are chances that the growth rate keeps negative in a certain time interval.  The Fourier amplitude takes very small values for the time interval. Inversely, even if the average value of the growth rate is negative, there are chances that the growth rate keeps positive in a certain time interval.  The Fourier amplitude takes non-small values for the time interval.  This is a mechanism that a random multiplicative process  induces an intermittent time sequence \cite{rf:13,rf:14,rf:15}.  When the amplitude of spatial fluctuations becomes large, the nonlinear term acts and the growth of the spatial fluctuations is suppressed. 
Figures 4(a) and (c) display time sequences of Fourier amplitudes 
$|\phi_k|$ with (a) $k=0.471$ and (c) $k=0.942$ at $D=0.2$. 
The average growth rate is positive for wave number $k=0.471$ and 
the average growth rate is negative for wave number $k=0.942$. For both wave numbers, intermittent time sequences appear.  Figures 4(b) and (d) display histograms of the probability distributions $P(|\phi_k|)$ of 
$|\phi_k|$ for (a) $k=0.471$ and (b) $k=0.942$.  The probability distributions obey 
approximately a power law for a rather large range of $k$, which is characteristic of multiplicative stochastic processes \cite{rf:14,rf:15}. The exponent of the power-law distribution is nearly -0.74 for $k=0.471$ and -0.93 for $k=0.942$.  The exponents change with wave numbers.  
Similar type intermittent behaviors are also observed for $D=2.5$. 

To summarize, we have found noise-induced synchronization for the modified Kuramoto-Sivashinsky equation.  The phase turbulence is suppressed by the common noise.  The mechanism for the noise-induced synchronization is clear in this model equation. That is, the stationary phase distribution of the uniform state is 
deformed by the noise and the uniform state is stabilized against the long-wavelength fluctuations. 
For $r=0$ as in the original Kuramoto-Sivashinsky equation, this type of noise-induced synchronization does not occur. 
The growth rate of the Fourier amplitude with the long wavenumber is decreased 
especially for an intermediate range of the noise sterngth. 
The average growth rate is positive for a finite interval of wave numbers, and the spatial profile becomes rather regular owing to the common noise.  The fluctuation of the growth rate 
induces an intermittent time sequence and a power-law distribution for the Fourier amplitude has been observed.

\end{document}